\documentclass[preprint,12pt]{elsarticle}

\usepackage{amssymb}
\usepackage{graphicx}
\usepackage{hyperref}

\begin{document}

\begin{frontmatter}


\title{Notes on remanent magnetization measurements in superconductors and hard ferromagnets}
\author[ISU]{Sergey L. Bud'ko\corref{cor1}}
\author[MSU]{Mingyu Xu}
\author[MSU]{Weiwei Xie}
\author[UCLA]{Chaowei Hu\fnref{fn1}}
\author[UCLA] {Ni Ni}
\author[ISU]{Paul C. Canfield}
\cortext[cor1]{Corresponding author. E-mail address: budko@ameslab.gov}
\fntext[fn1]{Currently at Department of Physics , University of Washington, Seattle, WA 98195, USA}
\affiliation[ISU]{organization={Ames National Laboratory and Department of Physics and Astronomy, Iowa State University },
             city={Ames},
             postcode={50011},
             state={Iowa},
             country={USA}}
\affiliation[MSU]{organization={Department of Chemistry, Michigan State University},
             city={East Lansing},
             postcode={48824},
             state={Michigan},
             country={USA}}
\affiliation[UCLA]{organization={Department of Physics and Astronomy and California NanoSystems Institute, University of California,
Los Angeles,},
             city={Los Angeles},
             postcode={90095},
             state={California},
             country={USA}}






\begin{abstract}
Data on zero applied field measurements of remanent magnetization and magnetic relaxation in a BCS superconductor LuNi$_2$B$_2$C and several hard ferromagnets are presented and compared. Apparent similarities and differences, in particular in Thermoremanent Magnetization (TRM) - like, Isothermal Remanent Magnetization (IRM) - like, and remanent magnetization measurements with zigzag temperature sweep measurements are outlined. It is discussed how these results could be relevant for the magnetization measurements in diamond anvil cells. 

\end{abstract}



\begin{keyword}

superconductors \sep pinning \sep trapped flux \sep ferromagnets \sep remanent magnetization



\end{keyword}

\end{frontmatter}


\section{Introduction}

Measurements of remanent magnetization in geophysical samples, \cite{cox61a} and spin glasses \cite{myd93a}  and of trapped flux magnetization in superconductors \cite{pip55a} have been known for decades. Trapped flux magnetization in superconductors, in paticular, had periods of renewed interest, either for potential applications, \cite{tom03a,fuc13a} or for identification and studies of new superconductors. \cite{mue87a,mos91a} More recently the measurements of remanent magnetization (trapped flux magnetization) were suggested \cite{hir21a}  and interpreted \cite{min23a} as a strong evidence of superconducting state in superhydrides under pressures exceeding a megabar.  In this case \cite{min23a} of particular importance was that the trapped flux measurements protocol (measurements are performed in the $H = 0$ applied field) allowed for the avoidance, or minimization, of significant contribution to the magnetic signal coming from the diamond anvil pressure cell (DAC). Although trapped flux measurements, especially when taken together with other experimental signatures such as zero electrical resistance, have been considered to be strong evidence of superconductivity, \cite{mue87a} there remains controversy regarding other possible ground states that could mimic the signatures of superconductivity in trapped flux measurements. \cite{hir22a}  In this paper we present experimental protocols and data to more clearly distinguish ferromagnetic transitions from superconducting ones.

The trapped flux magnetization measurements in DACs poised two questions: (a) can such measurements protocols allow us to extend the available pressure range for studies of superconducting properties of known superconductors to significantly higher pressures, and (b) can such measurements serve as a reliable way of identifying superconducting state in new, pressure stabilized, materials? The first question was answered affirmatively e.g. in Ref. \cite{huy25a} where self-field critical current density and base temperature flux creep rate as well as $T_c$ of CaKFe$_4$As$_4$ were measured in DAC up to $\sim 7.5$~GPa. The answer to the second question appears to be less straightforward. On one hand, the authors of Ref. \cite{mue87a} wrote "When we field cool the sample ... to low temperature and then switch off the field, a positive remanent magnetization is observed ... . This remanent magnetization results from flux trapping and it is a proof of superconductivity in its own right''. On the other hand, it was noted \cite{pro24a} that curves similar to the applied field dependence of the trapped flux magnetization in a superconductor can be obtained if similar measurements are performed on a ferromagnetic sample with domain walls pinning.

To further address this second question, in this work we present and compare experimental remanent magnetization data as a function of temperature and magnetizing field,  taken in zero-field-cooled (ZFC) and field-cooled (FC) protocols, derived TRM and IRM plots, remanent magnetization with zigzag temperature sweeps, and magnetization relaxation measurements taken at ambient pressure on an established type II superconductor and several ferromagnets with domain walls pinning and coercivity.

\section{Experimental}

The measurements were performed on superconducting LuNi$_2$$^{10}$B$_2$C single crystal, grown as described in Refs. \cite{xum94a,can01a} with the $^{10}$B isotope,  ferromagnetic single crystals of LaCrGe$_3$, \cite{lin13a,xum23a} SmCrGe$_3$, \cite{xum24a} MnBi$_8$Te$_{13}$ \cite{huc24a} and ferromagnetic polycrystalline sample of  Ce$_{3-x}$Mg$_x$Co$_9$, ($x \approx 0.6$). \cite{lam18a,lam20a} LuNi$_2$B$_2$C and MnBi$_8$Te$_{13}$ crystals are plate - like, with the $c$ - axis perpendicular to the plate, whereas LaCrGe$_3$ and SmCrGe$_3$ are rod - like with the $c$ - axis along the rod. 

The magnetic measurements were performed using a Quantum Design MPMS3 magnetometer. The samples were fixed on a MPMS3 semi-cylindrical sample using a small amount of GE-7031 varnish. For all four single crystals the applied magnetic field was parallel to the $c$ - crystallographic axis.  In the case of LuNi$_2$B$_2$C and MnBi$_8$Te$_{13}$ crystals a small L-shaped adapter made out of 0.05 mm thick copper foil was used (as described in Ref. \cite{bud23a}).

For temperature dependent remanent magnetization measurements we used the protocols similar to those in Ref. \cite{bud23a}. For convenience of the reader we outline the protocols here as well. Zero-field cooled (ZFC): the sample was cooled down from above the ordering temperature ($T_c$ or $T_C$) in $H = 0$; after the temperature was stabilized at the target base temperature, the magnetic field was increased to the target field, $H_M$; after a 0.5 min dwell time in the stable field, the magnetic field was decreased to $H = 0$; then, after another 0.5 min dwell time, $M(T)$ measurements on warming started. Field-cooled (FC): the sample was cooled down from above $T_c/T_C$ in the target field; after the temperature was stabilized at the target base temperature, the magnetic field was decreased to $H = 0$; then after 0.5 min dwell time, $M(T)$ measurements start. The linear mode ($H(t)$ - linear, no oscillations) was used to change the magnetic field. For the remanent magnetization measurements in the ZFC protocol with $H_M \leq 100$ Oe,  the samples were cooled in $H = 0$ from above $T_c/T_C$, after the demagnetization procedure for the superconducting magnet of MPMS3 was performed.

For the remanent magnetization measurements with zigzag temperature sweeps, the FC protocol with the applied field higher than the saturation field for the respective sample was used.

The magnetic relaxation measurements were performed in the FC protocol, again with the applied field value above the saturation field.

\section{Results}

\subsection{LuNi$_2$B$_2$C superconductor}

\begin{figure*}
  \includegraphics[width=1.0\textwidth]{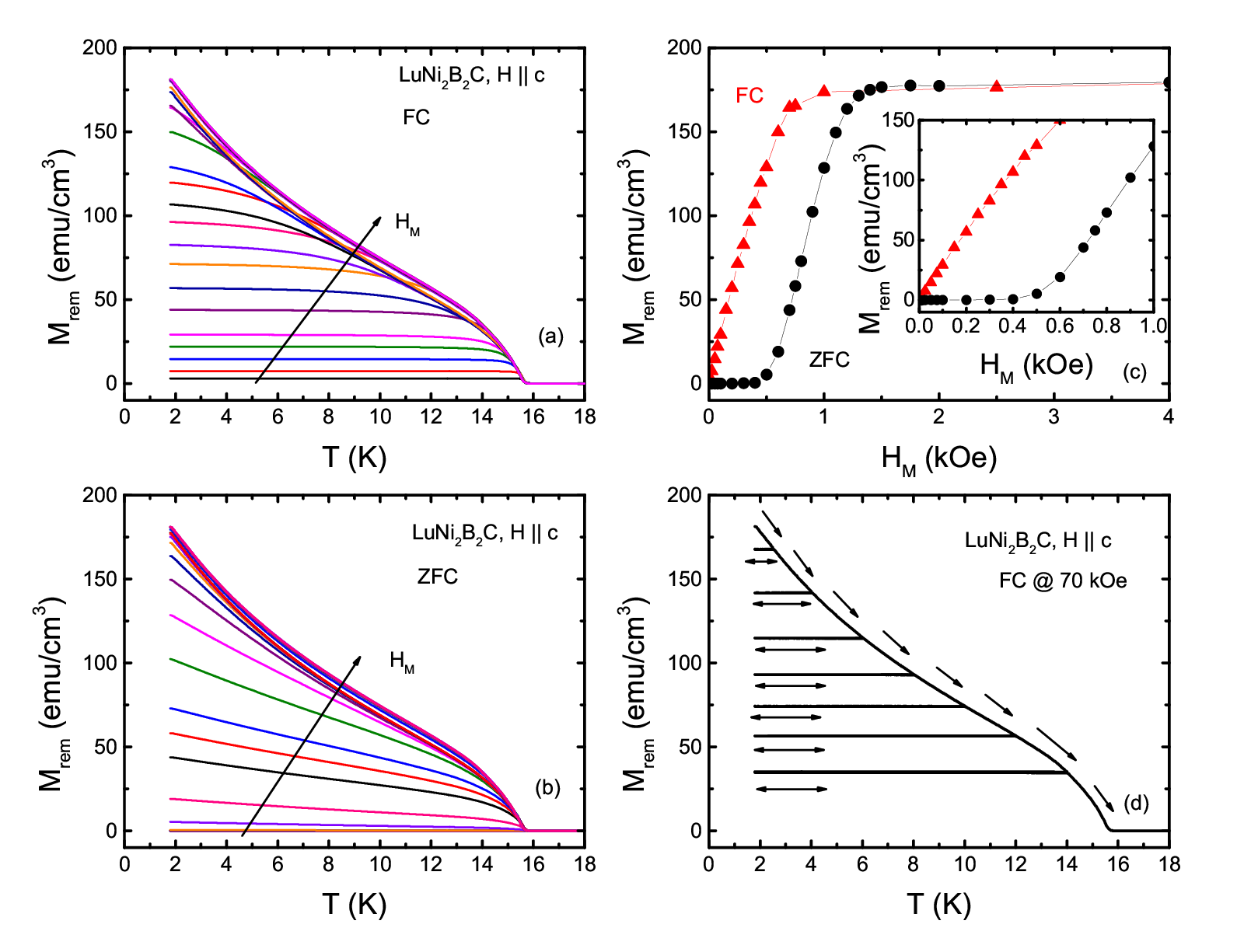}
\caption{(color online) (a) Temperature - dependent remanent (trapped flux) magnetization of LuNi$_2$B$_2$C obtained in FC protocol, arrow points to the direction of increase of the target fields $H_M \| c$, the values of $H_M$ in kOe are 0.01, 0.025, 0.05, 0.075, 0.1, 0.15, 0.2, 0.25, 0.3, 0.35, 0.4, 0.45, 0.5, 0.6, 0.7, 0.75, 1, 2.5, 5, 7.5, 10, 20, 40, 60, 70. (b) the same as (a) with the data obtained in ZFC protocol with $H_M$ values of 0.01, 0.025, 0.05, 0.075, 0.1, 0.2, 0.3, 0.4, 0.5, 0.6, 0.7, 0.75, 0.8, 0.9, 1, 1.1, 1.2, 1.3, 1.4, 1.5, 1.75, 2, 4, 6, 8, 10, 20, 40, 60, 70, kOe. (c) Remanent (trapped flux) magnetization at $T = 1.8$~K as a function of target field $H_M$ in the ZFC and FC experiments (Equivalent to TRM and IRM curves). The inset shows an enlarged, low field, part of the data. (d) Temperature - dependent remanent magnetization measured after FC in 70~kOe field with zigzag (shown by arrows) temperature sweep. }
\label{F1}       
\end{figure*}

LuNi$_2$B$_2$C is a well established superconductor with the critical temperature $T_c \approx 16$~K. \cite{cav94a,tak94a,che99a,sch01a} Figures \ref{F1}(a),(b) present temperature - dependent remanent (trapped flux) magnetization measured using FC and ZFC protocols. Fig. \ref{F1}(c) shows  remanent (trapped flux) magnetization at $T = 1.8$~K as a function of target field $H_M$. It is noteworthy, but not surprising, that these data are qualitatively similar to the data on other known ambient pressure superconductors, like MgB$_2$, CaKFe$_4$As$_4$, and CaK(Fe$_{0.983}$Mn$_{0.017}$)$_4$As$_4$ \cite{bud23a} as well as the data for  H$_3$S and LaH$_{10}$ at high pressures. \cite{min23a} Fig. \ref{F1}(d) shows temperature - dependent remanent magnetization measured after FC in 70~kOe field with zigzag temperature sweep. The parts of the data corresponding to cooling from an intermediate temperature $T_i < T_c$ and subsequent warming back to $T_i$ are horizontal, without any slope. Similar behavior was observed in other superconductors ranging from Sn - Pb solders \cite{miz24a} to syperhydride superconductorss at high pressure. \cite{min23a}

\begin{figure*}
  \includegraphics[width=0.8\textwidth]{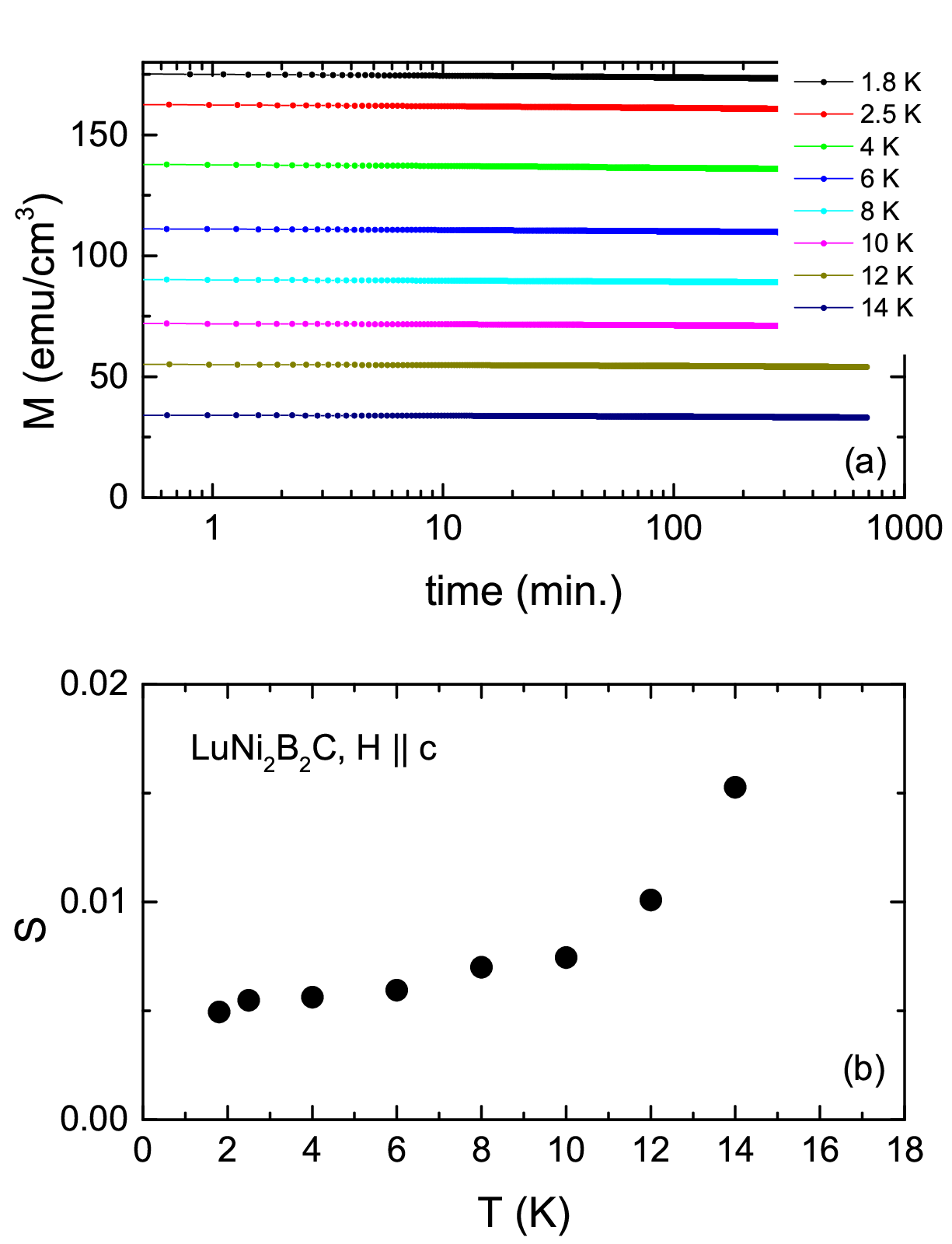}
\caption{(Color online) (a) Self-field magnetic relaxation (flux creep) in LuNi$_2$B$_2$C measured at different temperatures after 70 kOe FC. (b) Flux creep rate, $S = - 1/M(0) \times dM/d ln(t)$ as a function of  temperature.}
\label{F2}       
\end{figure*}

Fig. \ref{F2} presents self-field magnetic relaxation (flux creep) measurements in LuNi$_2$B$_2$C at several temperatures after FC in 70~kOe field as well as the temperature dependence of the flux creep rate. The observed increase of the flux creep rate on increase of temperature is consistent with a simple picture of decrease of pinning on approaching $T_c$. The data in Fig. \ref{F2}(b) are qualitatively similar to those measured in CaKFe$_4$As$_4$ and MgB$_2$ single crystals \cite{bud24a} and actually are close in values to those of MgB$_2$ if plotted vs reduced temperature $T/T_c$.

\subsection{Ferromagnets}

\subsubsection{LaCrGe$_3$ crystal}

\begin{figure*}
  \includegraphics[width=0.75\textwidth]{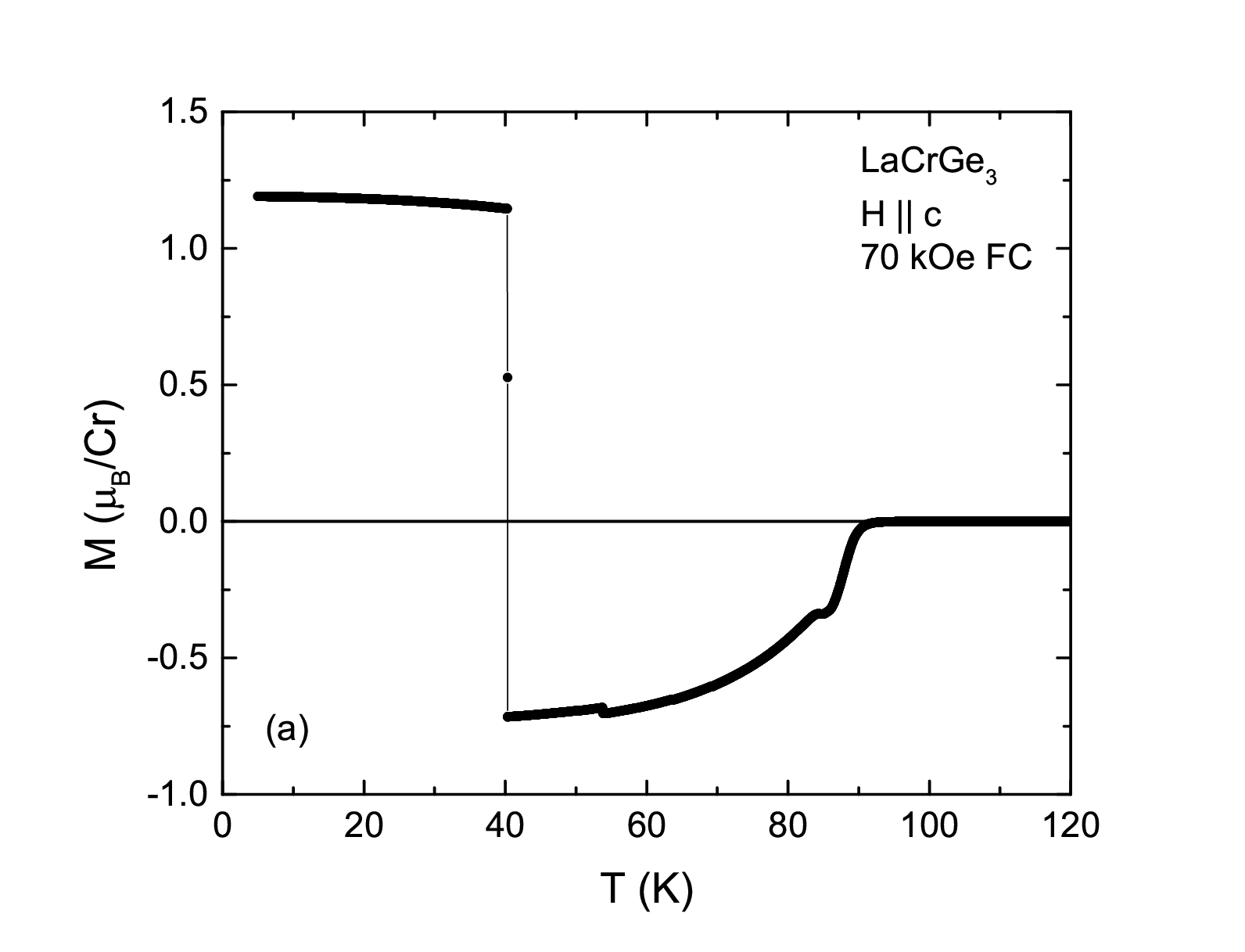}
  \includegraphics[width=0.75\textwidth]{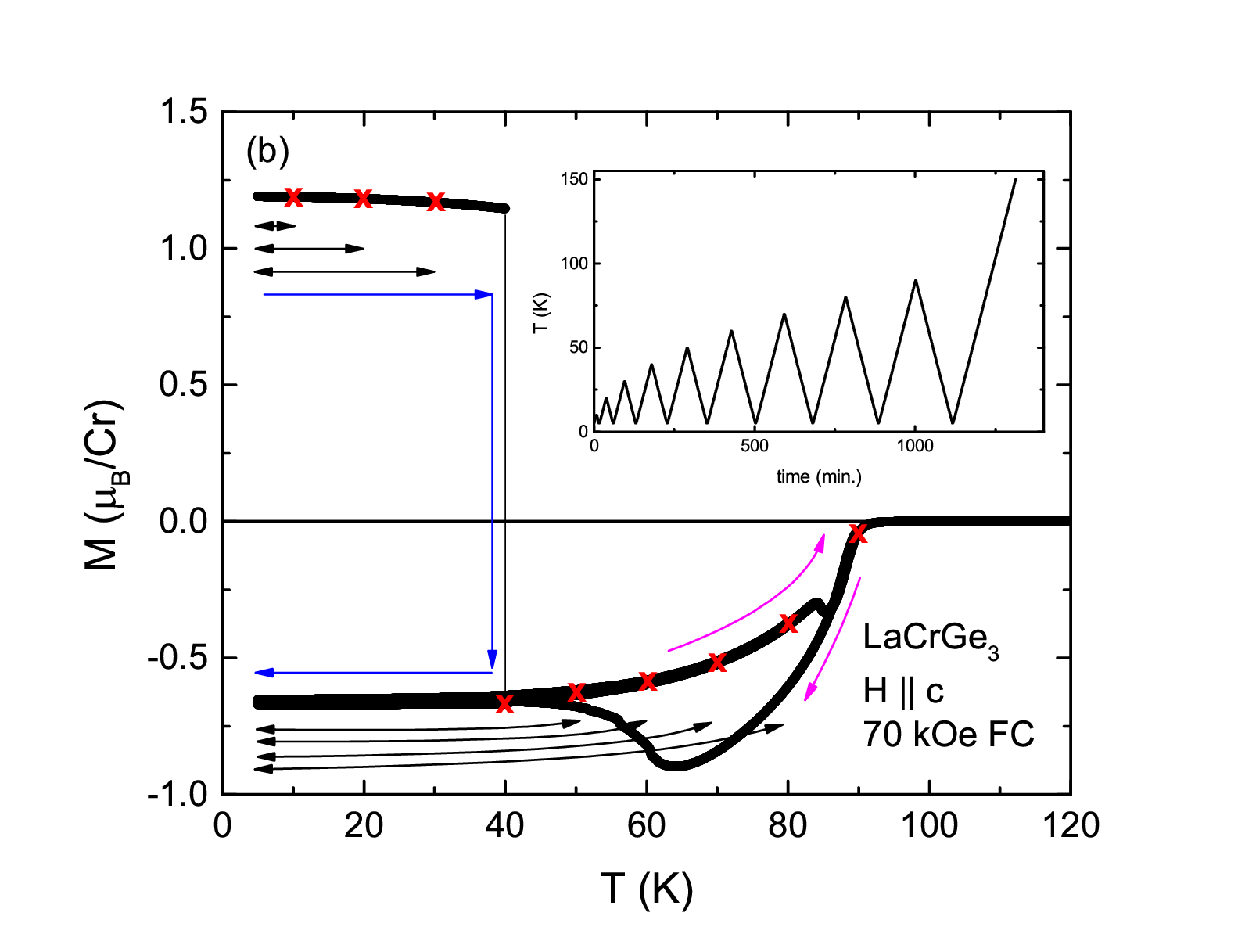}
\caption{(Color online) (a) Temperature dependent remanent magnetization data for LaCrGe$_3$ single crystal taken uzing70~kOe  FC protocol ($H \| c$). (b) Temperature dependent remanent magnetization measurements after FC in 70~kOe field with zigzag (shown by arrows) temperature sweep. Intermediate temperatures $T_i$ are shown by red "X". Except for intermediate temperatures $T_i = 40$~K and 90~K (blue and magenta arrows, respectively) the up and down sweeps are reversible. Inset shows temperature vs time protocol for this measurement.}
\label{F3}       
\end{figure*}

LaCrGe$_3$ orders ferromagnetically at $T_C \approx 86$~K. \cite{lin13a,xum23a} A 70~kOe ZFC temperature dependent remanent magnetization data for LaCrGe$_3$ are shown in Fig. \ref{F3}(a). The magnetization sign reversal at $\sim 40$~K is highly unusual and was studied in detail in Ref. \cite{xum23a}. LaCrGe$_3$ was found to have remarkable low temperature coercivity that is temperature dependent: it drops to zero in the 40 - 55~K region and reappears in the 70 - 85 K regions. This region of zero coercivity at the intermediate temperatures below $T_C$ was linked to the magnetization reversal. \cite{xum23a} Fig. \ref{F3}(b) presents the temperature dependent remanent magnetization measured using zigzag temperature sweep. We will not discuss LaCrGe$_3$ further here, other than noting that its remanent magnetization is distinctly different from that observed in superconductors.

\subsubsection{SmCrGe$_3$ crystal}

\begin{figure*}
  \includegraphics[width=1.0\textwidth]{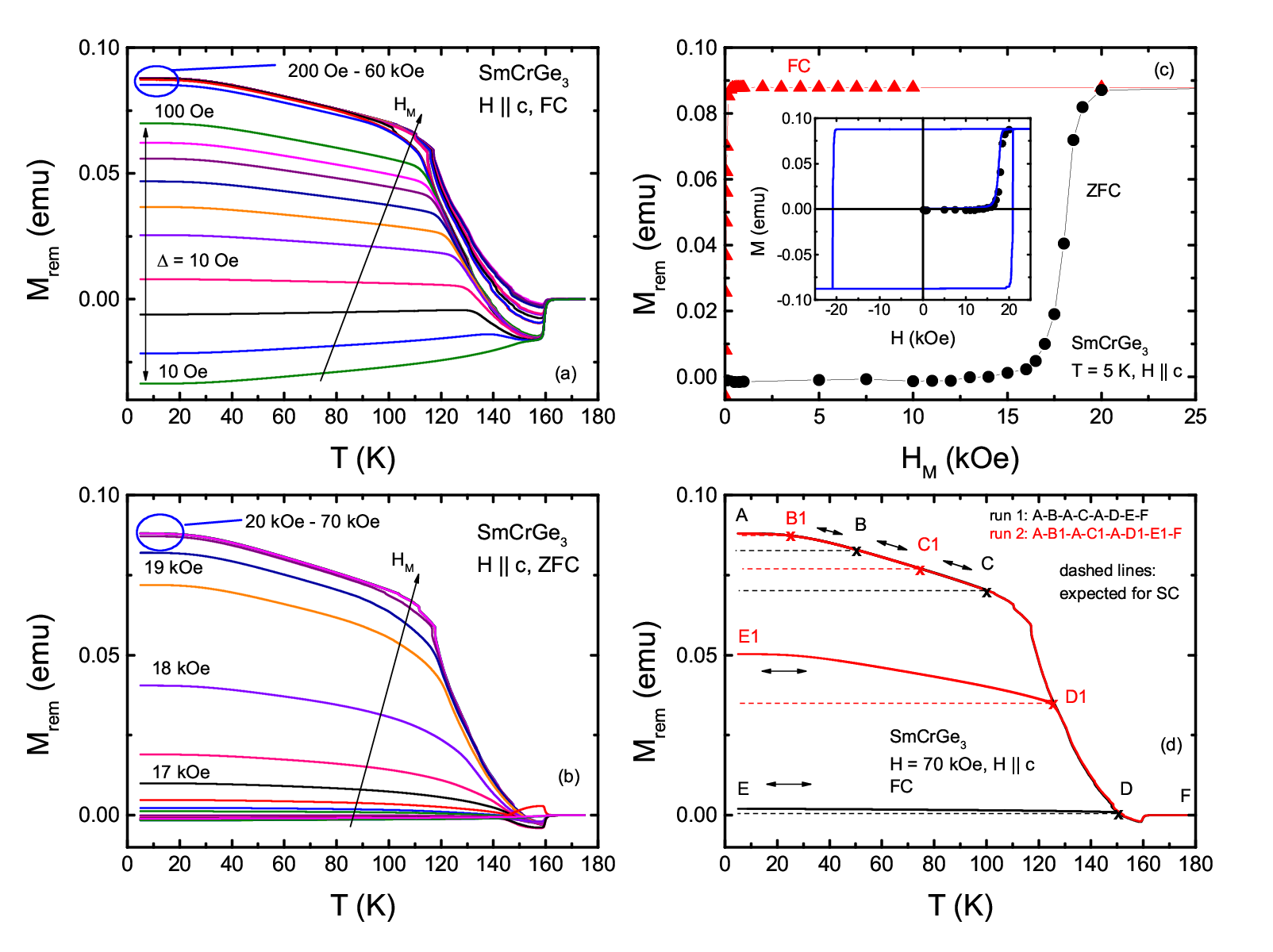}
\caption{(color online) (a) Temperature - dependent remanent magnetization of SmCrGe$_3$ obtained in FC protocol, arrow points to the direction of increase of the target fields $H_M \| c$, the values of $H_M$ in kOe are 0.01, 0.02, 0.03, 0.04, 0.05, 0.06, 0.07, 0.08, 0.09, 0.1,  0.2, 0.3, 0.4, 0.5, 0.6, 0.7, 0.8, 0.9, 1, 2, 3, 4, 5, 6, 7, 8, 9, 10, 20, 30, 40, 50, 60. (b) the same as (a) with the data obtained in ZFC protocol with $H_M$ values of 0.1, 0.5, 0.75, 1, 5, 7.5, 10, 11, 12, 13, 14, 15, 16, 16.5, 17, 17.5, 18, 18.5, 19, 20, 30, 40, 60, 70, kOe. (c) Remanent magnetization at $T = 5$~K as a function of target field $H_M$ in the ZFC and FC experiments The inset shows ZFC part of the data (black line plus symbol) and 5 - quadrants magnetization loop (blue line) measured at $T = 5$~K. (d) Two runs of the temperature - dependent remanent magnetization measurements after FC in 70~kOe field with zigzag (shown by arrows) temperature sweep. Dashed horizontal lines show behavior expected for a superconductor [cf. Fig. \ref{F1}(d)].}
\label{F4}       
\end{figure*}

SmCrGe$_3$ is a close cousin of LaCrGe$_3$, a ferromagnet with $T_C \approx 160$~K.  It has large magnetocrystalline anisotropy and monotonic in temperature coercivity without apparent anomalies. \cite{xum24a} Figures \ref{F4}(a),(b) present temperature - dependent remanent magnetization measured using FC and ZFC protocols, whereas Fig. \ref{F4}(c) shows  remanent magnetization at $T = 5$~K as a function of the target field $H_M$. It is noteworthy that in particular in FC but also in ZFC measurements at low values of $H_M$ the measured remanent magnetization is negative [Figs. \ref{F4}(a),(b)]. In our opinion this is an artefact caused by magnetic flux trapped in the superconducting magnet of the magnetometer (see some discussion in Ref. \cite{bud23a} for the case of measurements on superconductors). Note, that this artefact, in principle, can be to a large extent removed by applying small opposite sign field to balance remanent field in the magnet (Fig. A8 in Ref. \cite{bud23a}). Those procedures are time and resources consuming. 

FC remanent magnetization [Fig. \ref{F4}(c)] reaches saturation at a relatively low field of $\sim 0.3$~kOe, whereas ZFC remanent magnetization is effectively zero up to $\sim 12$~kOe, and the saturation is reached at $\sim 20$~kOe, way higher than for FC. Not surprisingly, the ZFC remanent magnetization data at 5~K follow the "virgin" $M(H)$ curve measured at the same temperature [inset to Fig. \ref{F4}(c)]. 

Fig. \ref{F4}(d) shows two runs of temperature - dependent remanent magnetization measurements after FC in 70~kOe field with zigzag temperature sweeps. The intermediate, turn-around, temperatures, $T_i$, were different in these runs. For $T_i \lesssim 110$~K the behavior is reversible: $M_{rem}(T)$ is the same on cooling and warming and it follows saturated $M_{rem}(T)$ curves as in Figs. \ref{F4}(a),(b). For higher temperatures there is a clear difference: the cooling curve [e.g. D1-E1 in fig. \ref{F4}(d)]  is different from initial cooling (A-D1), but then warming E1-D1 segment is the same as the previous D1-E1 cooling, and on further warming (D1-F) the behavior follows saturated $M_{rem}(T)$ curve. This behavior can be rationalized as follows: below $\sim 110$~K either the sample behaves as a single domain or the domains dynamics is reversible. Above $\sim 110$~K the regime is multi-domain, domains are (partially) misaligned and this process is not reversible. On cooling from an intermediate temperature (e.g. D1-E1 curve) the measured $M_{rem}(T)$ increases simply due to increase of the magnetic moment. For higher intermediate temperatures (e.g. D-E curve) the domains are almost completely misaligned and the increase of $M_{rem}(T)$ on cooling is present but rather small. This behavior is qualitatively different from that observed in superconductors with pinning (see above the case of LuNi$_2$B$_2$C), as in superconductors once the vortices exit the sample on warming they do not re-enter on cooling and respective parts of the $M_{rem}(T)$ curve are horizontal as shown by dashed lines in Fig. \ref{F4}(d).

\begin{figure*}
  \includegraphics[width=0.9\textwidth]{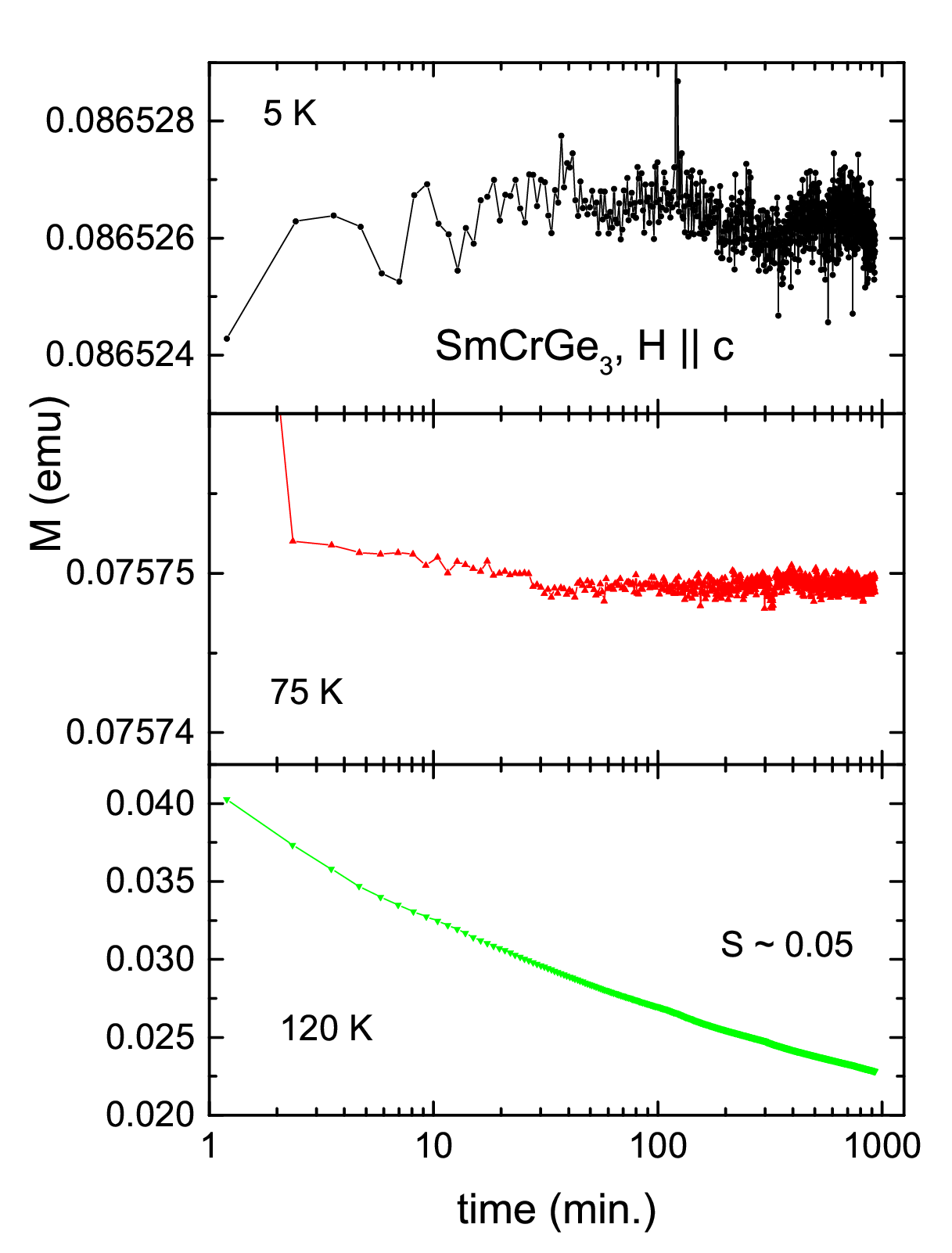}
\caption{(Color online) Magnetic relaxation in SmCrGe$_3$ measured at three different temperatures after 70 kOe FC and decreasing magnetic field to zero. A rough estiate of the relaxtion rate, $S = - 1/M(0) \times dM/d ln(t)$, is listed for the 120~K data (lower panel).}
\label{F5}       
\end{figure*}

Time dependence of the self-field remanent magnetic moment of SmCrGe$_3$ measured at three different temperatures below $T_C$ is presented in Fig. \ref{F5}. For 5~K and 75~K ($T/T_C \approx 0.03$ and 0.5, respectively, the single-domain region) the changes in $M_{rem}$ over $\sim 1000$~min. are in tens of ppm, for 120~K ($T/T_C = 0.75$, multi-domain region) the relaxation is close to exponential (with minor deviations), with the relaxation rate, $S = - 1/M(0) \times dM/d ln(t)$, roughly 0.05.

\subsubsection{MnBi$_8$Te$_{13}$ crystal}

\begin{figure*}
  \includegraphics[width=1.0\textwidth]{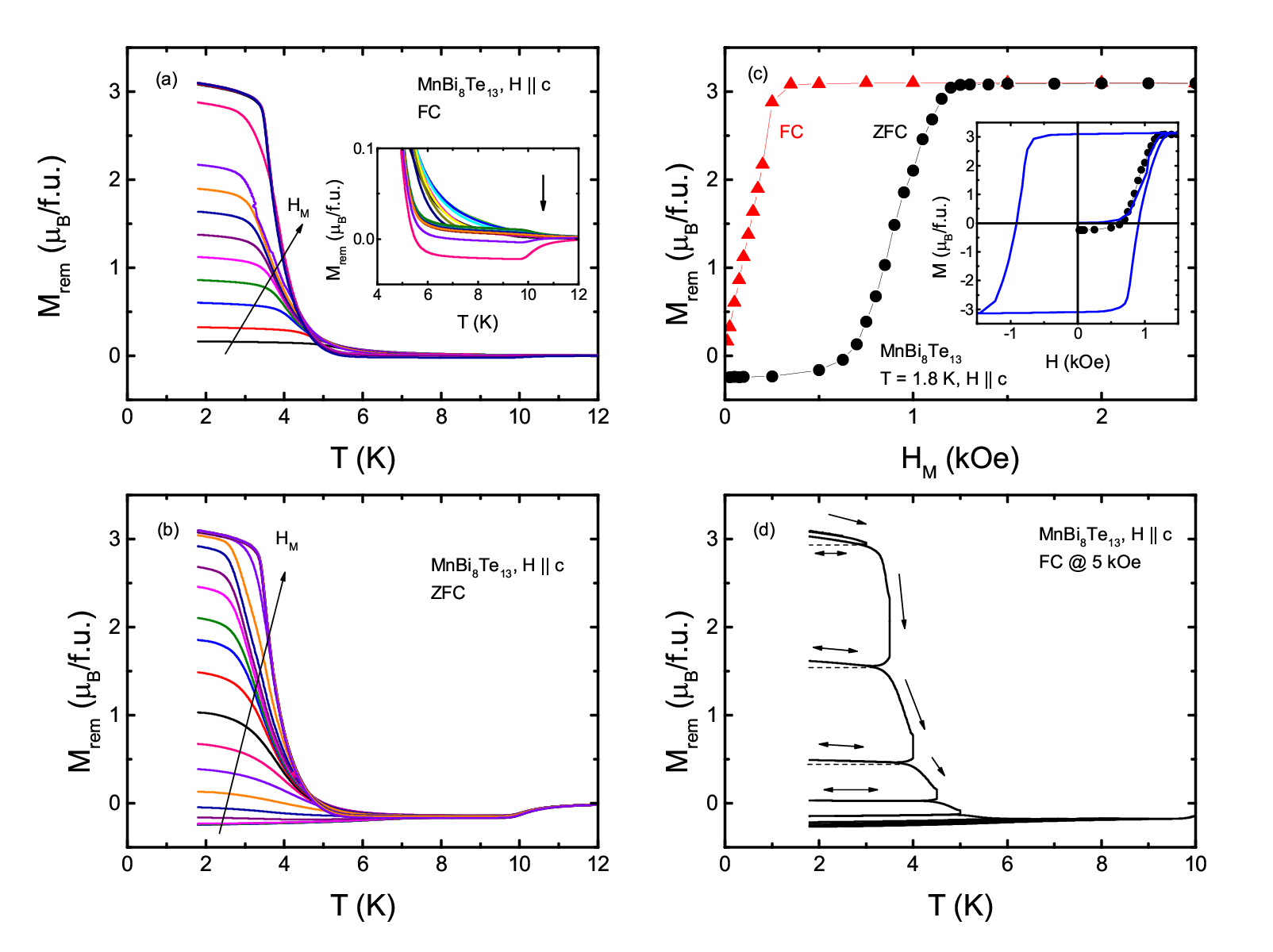}
\caption{(color online) (a) Temperature - dependent remanent magnetization of MnBi$_8$Te$_{13}$ obtained in FC protocol, arrow points to the direction of increase of the target fields $H_M \| c$, the values of $H_M$ in kOe are 0.0125, 0.025, 0.05, 0.075, 0.1, 0.125, 0.15, 0.175 0.2, 0.25, 0.35, 0.5, 0.75, 1, 1.5, 2, 2.5. The inset shows $M_{rem}(T)$ close to zero to emphasize an artefact of negative values for low $H_M$. The vertical arrow corresponds to $T_C$. (b) the same as (a) with the data obtained in ZFC protocol with $H_M$ values of 0.025, 0.05, 0.075, 0.1, 0.25, 0.5,0.625, 0.7, 0.75, 0.8, 0.85, 0.9, 0.95, 1, 1.05, 1.1, 1.15, 1.2, 1.25, 1.3, 1.4, 1.5, 1.75, 2, 2.25, 2.5, 2.75, 3, 4 kOe. (c) Remanent magnetization at $T = 1.8$~K as a function of target field $H_M$ in the ZFC and FC experiments The inset shows ZFC part of the data (black line plus symbol) and 5 - quadrants magnetization loop (blue line) measured at $T = 1.8$~K. (d) Temperature - dependent remanent magnetization measurements after FC in 5~kOe field with zigzag (shown by arrows) temperature sweep. Dashed horizontal lines show behavior expected for a superconductor [cf. Fig. \ref{F1}(d)].}
\label{F6}       
\end{figure*}

MnBi$_8$Te$_{13}$ belongs to a family of MnBi$_{2n}$Te$_{3n+1}$ intrinsic magnetic topological insulators. \cite{huc24a} It is a ferromagnet with $T_C \sim 10$~K. It was reported to have relatively narrow hysteresis loops and soft magnetic domains. \cite{huc22a} One of the features seen in the temperature - dependent remanent magnetization data measured using FC and ZFC protocols [Figs. \ref{F6}(a),(b)], is that the remanent magnetization becomes very small already at $\sim 5$~K ($\sim 0.5 T_C$), yet the value of $T_C$ can be inferred from these data. The reason for this is a combunation of narrow hysteresis, decreasing with temperature, and the demagnetizing field skewing effect \cite{sko99a} (see also \ref{A}). As in other materials, the artefact of negative $M_{rem}$ for low $H_M$ is observed.

FC remanent magnetization [Fig. \ref{F6}(c)] reaches saruration at $\sim 0.3$~kOe, whereas ZFC remanent magnetization is approximately flat up to $\sim 0.2$~kOe, and the saturation is reached at $\sim 1.35$~kOe, factor of $\sim 4$ higher than for FC. The ZFC remanent magnetization data at 1.8~K follow the "virgin" $M(H)$ curve measured at the same temperature [inset to Fig. \ref{F6}(c)]. 

Fig. \ref{F6}(d) shows temperature - dependent remanent magnetization measurements after FC in 5~kOe field with zigzag temperature sweeps. Similar to SmCrGe$_3$, although less pronounced, $M_{rem}(T)$ increases on cooling from an intermediate temperature, is reversible almot to the turn-over temperature, and then approximately follows the saturation $M_{rem}(T)$ on further warming.

\begin{figure*}
  \includegraphics[width=0.9\textwidth]{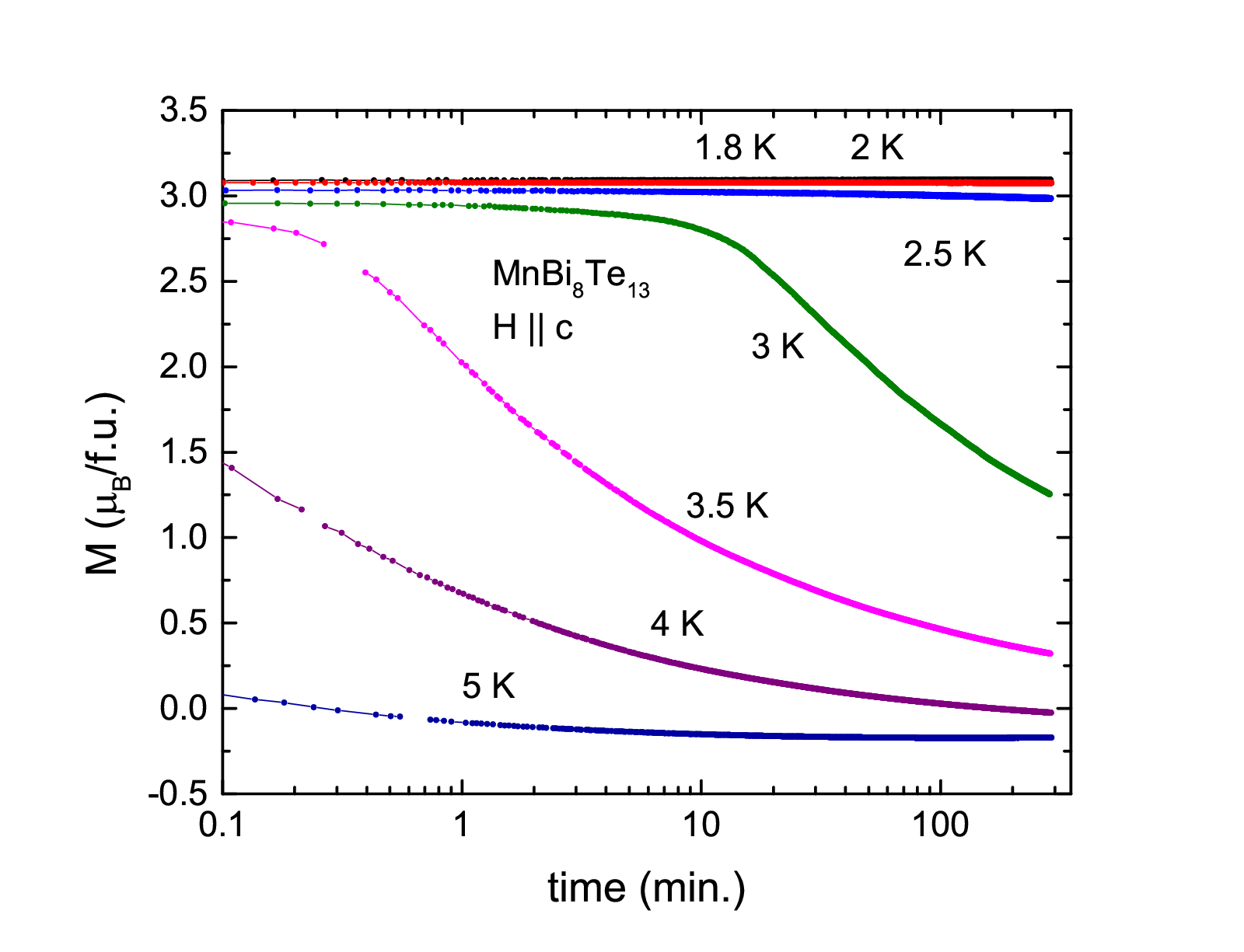}
\caption{(Color online) Magnetic relaxation in SmCrGe$_3$ measured at select temperatures after 5 kOe FC and decreasing the magnetic field to zero.}
\label{F7}       
\end{figure*}

The magnetic relaxation behavior (Fig. \ref{F7}) is rather unusual for 3 and 3.5~K. The data are consistent with those reported in Ref. \cite{huc22a} where the meaurements were only $\sim 8.3$ minutes long.

\subsubsection{Polycrystal of Ce$_{3-x}$Mg$_x$Co$_9$ with $x \sim 0.6$}

\begin{figure*}
  \includegraphics[width=1.0\textwidth]{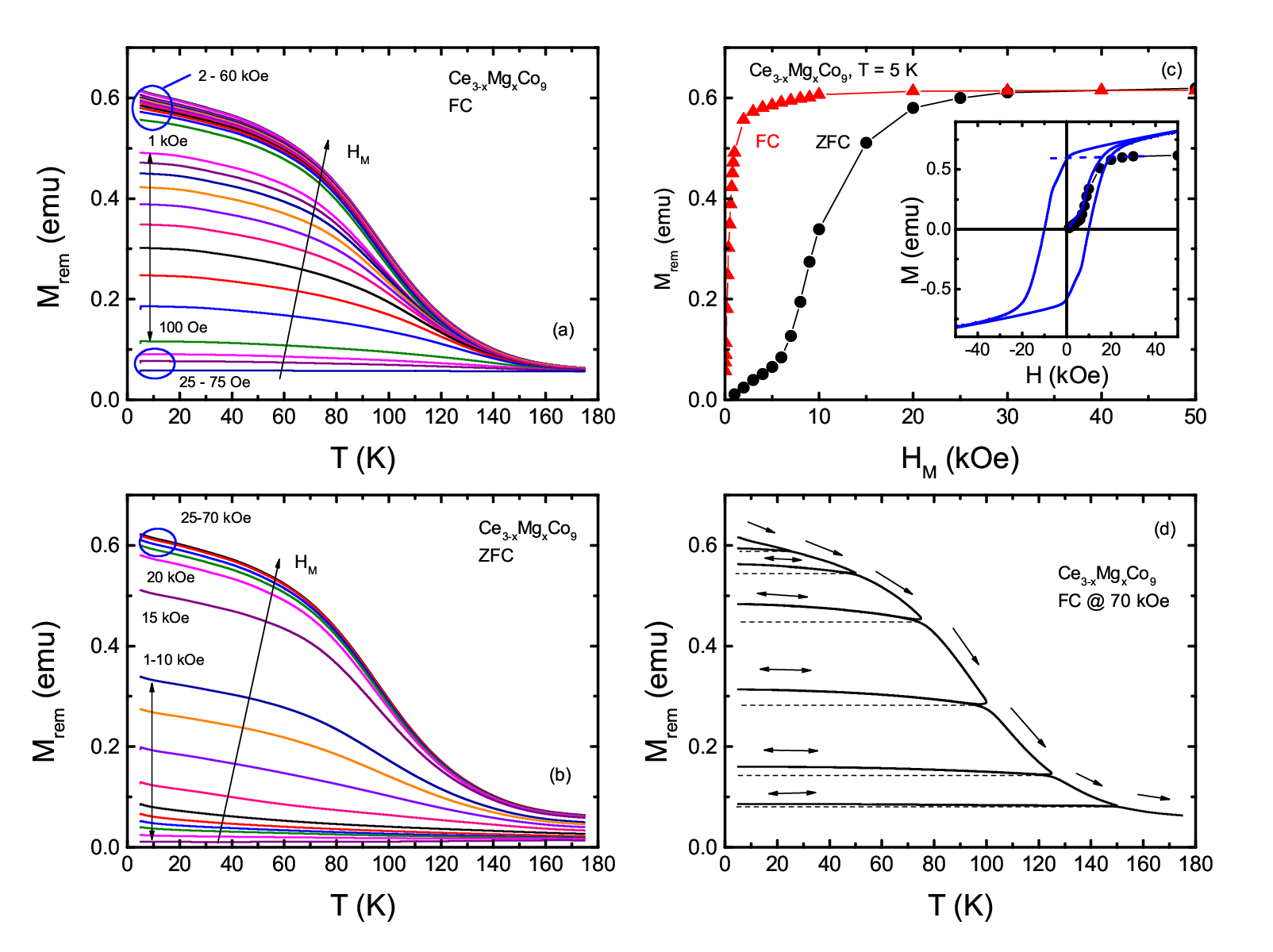}
\caption{(color online) (a) Temperature - dependent remanent magnetization of Ce$_{3-x}$Mg$_x$Co$_9$, $x \sim 0.6$, obtained in FC protocol, arrow points to the direction of increase of the target fields $H_M \| c$, the values of $H_M$ in kOe are 0.025, 0.05, 0.075, 0.1, 0.2, 0.3, 0.4, 0.5, 0.6, 0.7, 0.8, 0.9, 1, 2, 3, 4, 5, 6, 7, 8, 9, 10, 20, 30, 40, 50, 60. (b) the same as (a) with the data obtained in ZFC protocol with $H_M$ values of 1, 2, 3, 4, 5, 6, 7, 8, 9, 10, 15, 20, 25, 30, 50, 70 kOe. (c) Remanent magnetization at $T = 5$~K as a function of target field $H_M$ in the ZFC and FC experiments The inset shows ZFC part of the data (black line plus symbol) and 5 - quadrants magnetization loop (blue line) measured at $T = 5$~K. Horizontal dashed line corresponds to remanent magnetization from the $M(H)$ loop.(d) Temperature - dependent remanent magnetization measurements after FC in 70~kOe field with zigzag (shown by arrows) temperature sweep. Dashed horizontal lines show behavior expected for a superconductor [cf. Fig. \ref{F1}(d)].}
\label{F8}       
\end{figure*}

This sample is a part of Ce$_{3-x}$Mg$_x$Co$_9$ series studied in Refs.\cite{lam18a,lam20a}. It is a ferromagnet with $T_C \sim 140$~K. As other polycrystalline samples in this family, it probably has some distribution of the Mg concentrtion $x$. Fig. \ref{F8} presents the same datasets as were shown for SmCrGe$_3$ and MnBi$_{2n}$Te$_{3n+1}$ in figures \ref{F4} and \ref{F6} respectively. Qualitatively, these three datasets are similar. As above, the saturation fields for remanent magnetization measurements in FC and ZFC protocols are significantly (by factor of $\sim 3$) different. $M_{rem}(H_M)$ follows the virgin part of the $M(H)$ loop, although $M(H)$ does not saturates completely in these fields. The parts of the $M_{rem}(T)$ curve corresponding to cooling / warming at intermediate turn-around temperatures have clear negative slope ($M_{rem}(T)$ increases on warming).

\begin{figure*}
  \includegraphics[width=0.8\textwidth]{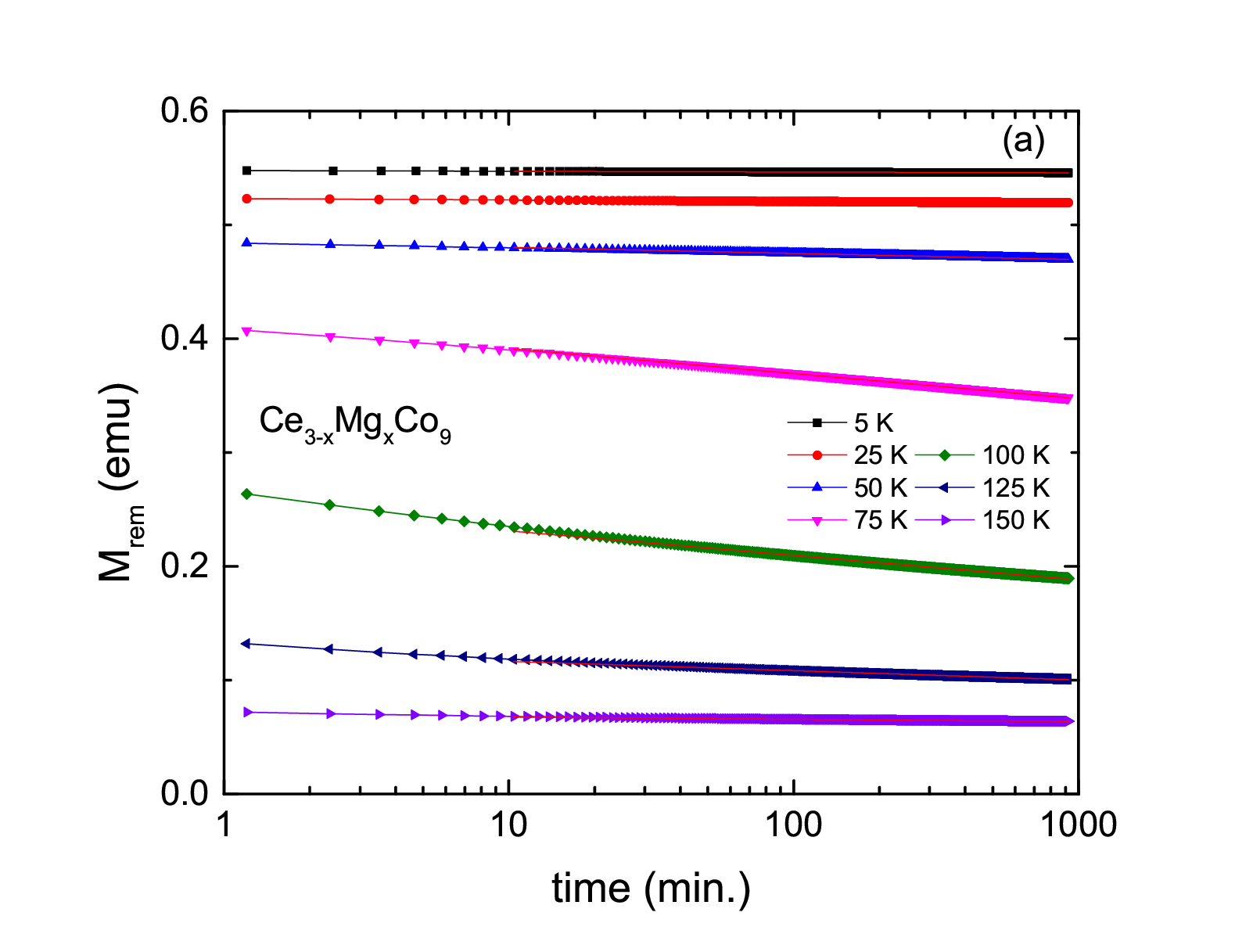}
  \includegraphics[width=0.8\textwidth]{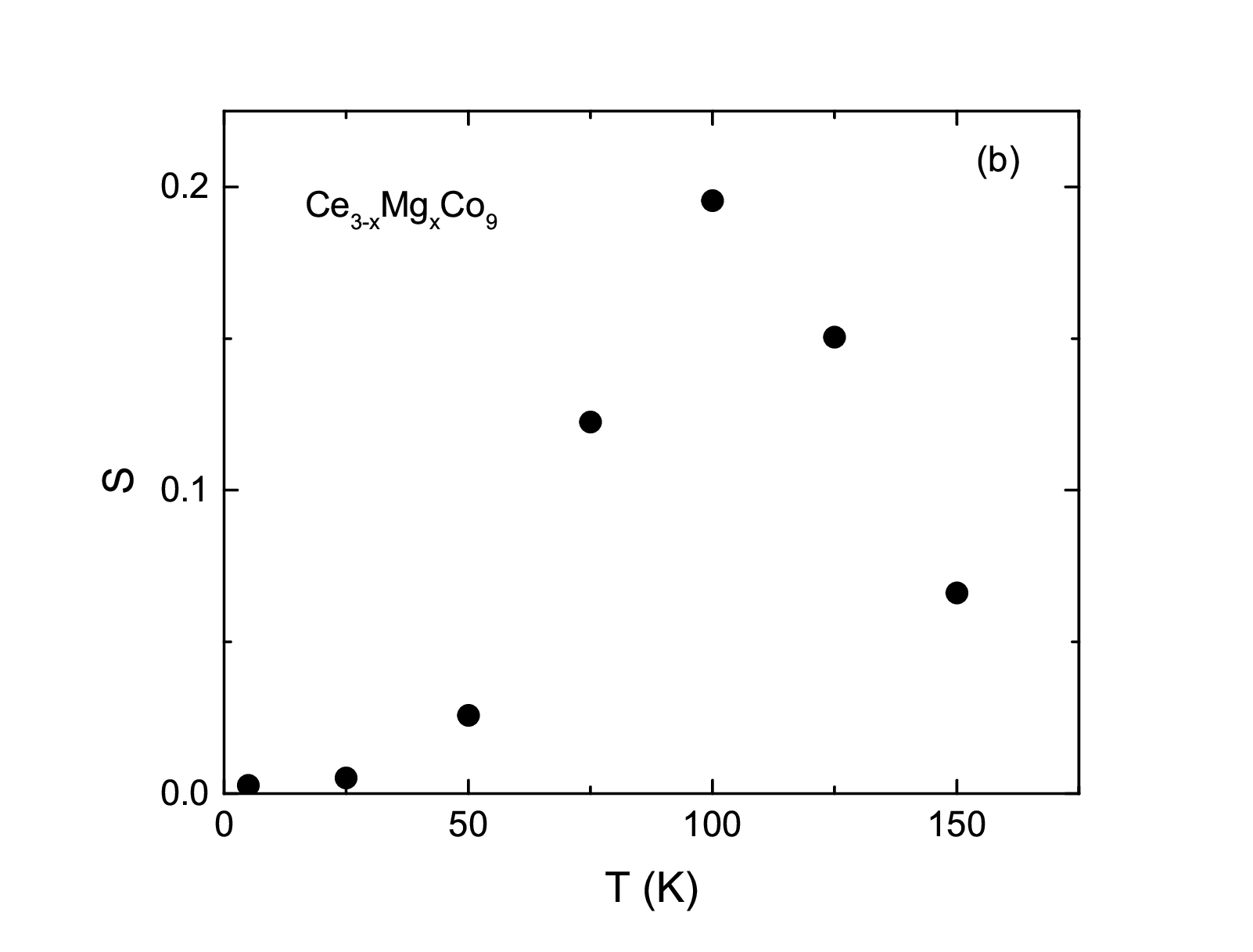}
\caption{(Color online) (a) Magnetic relaxation in Ce$_{3-x}$Mg$_x$Co$_9$, $x \sim 0.6$, measured at select temperatures after 70 kOe FC and decreasing the magnetic field to zero. (b) Magnetic relaxation rate in zero applied field plotted as a function of temperature.}
\label{F9}       
\end{figure*}

Magnetic relaxation curves measured at different temperatures in zero field after cooling down from above $T_C$ in 70~kOe  are shown in Fig. \ref{F9}(a). The relaxation rate $S$ obtained from these data is plotted in Fig. Fig. \ref{F9}(b) as a function of temperature. It has a maximum at $\sim 100$~K, and some relaxation is measured even at 150~K, above the $T_C$ of the majority phase of the sample.

\section{Discussion and Summary}

At first glance, qualitatively, the data presented here for the experimental remanent magnetization as a function of temperature and magnetizing field,  taken in ZFC FC protocols, derived TRM and IRM plots, remanent magnetization with zigzag temperature sweeps, and magnetization relaxation measurements, for type II superconductor with pinning and several ferromagnets with hysteresis, look similar. In the course of this work we emphasize the importance of identical protocols and conditions and metodological rigor if one wants to compare these two classes of materials. One needs to look in details to observe potential differences. 

Remanent  (trapped flux) magnetization in type II superconductors is usually described well by Bean's theory \cite{bea62a,bea64a} with some further corrections due to the shape (demagnetization factor) of the sample, pinning mechanisms, etc. Not surprisingly, such data for LuNi$_2$B$_2$C in this work are similar to the publshed results for other superconductors. \cite{min23a,bud24a}. In particular, (i) $M_{rem}(H_M)$ curve in ZFC protocol necessarily has, even if small, part with $M_{rem} = 0$ for $H < H_p$, where $H_p$ is the penetration field of the first vortex into the sample; (ii) in $M_{rem}(H_M)$ plots (or TRM and IRM plots) the saturation field for ZFC protocol ($H_{sat}^{ZFC}$) approximately equals to FC protocol saturation field ($H_{sat}^{FC}$) times two, with addition of  $H_p$, $H_{sat}^{ZFC} = 2 H_{sat}^{FC} + H_p$, and since $H_p$ is often rather small, $H_{sat}^{ZFC} \sim 2 H_{sat}^{FC}$ [as seen in Fig. \ref{F1}(a)] ; (iii) in remanent magnetization with zigzag temperature sweeps measirements, the $M_{rem}(T)$ parts of the data corresponding to cooling / warming at intermediate turn-around temperatures are horizontal, since once the vortices exit the sample on warming they do not re-enter on cooling. 

In ferromagnets, on the other hand, (i) the $M_{rem}(H_M)$ in ZFC protocol follows the virgin part of the $M(H)$, so it does not necessarily have  $M_{rem} = 0$, although, as in the examples of SmCrGe$_3$ and MnBi$_8$Te$_{13}$, very similar behavior might be observed; (ii) there is no expectation of any relation between  $H_{sat}^{ZFC}$ and $H_{sat}^{FC}$, except  $H_{sat}^{ZFC} > H_{sat}^{FC}$ (we observed $H_{sat}^{ZFC} / H_{sat}^{FC}$ between $\sim 3$ and $\sim 30$ in the FM examples above); (iii) since the magnetic moment in ferromagnets increase on cooling, the $M_{rem}(T)$ in measurements of the remanent magnetization with zigzag temperature sweeps should (and clearly do) have negative slopes for the data corresponding to intermediate turn-over temperature points.

Note that a protocol for thermomagnetic hysteresis measurements was suggested in the literature \cite{esq17a} as an experimental way to distinguish between superconducting and ferromagnetic behavior. This protocol should be considered for further studies and potential implementation.  On a different note: in this study we performed measurements on single crystals and a bulk polycrystalline sample thus avoiding potential complications related to granular nature of the samples. \cite{deu21a,gok20a} Detailed studies of the effects of granularity including particle size and connectivity require additional, separate study. 

The differences between superconductors and ferromagnets outlined above could be somewhat subtle and might be difficult to see unambiguously in the case of not very high signal-to-noise measurements, inhomogeneous samples, or samples with ill-defined geometry. The conclusions would be more solid if low field ZFC $M(T)$  ({\it decrease} of $M(T)$ at $T_c$ for superconductors vs. {\it increase} at $T_C$ for ferromagnets) or resistance / resistivity, including in magnetic field ({\it instrumental zero} below $T_c$ and {\it decrease} of the apparent transition temperature in magnetic field for superconductors vs. only {\it some decrease} of resistance below $T_C$ and {\it increase} of the apparent transition temperature in magnetic field for ferromagnets) can be reliably measured.

In summary, the examples above show a clear potential of nominal zero applied field magnetization measurements for identification and further studies of superconductors with pinning and hard ferromagnets under pressure in diamond anvil cells. Such studies already exist. \cite{min23a,huy25a,bud24b,che25a,huy25b}  The observation of trapped flux magnetization in coincidence with other experimental signatures of superconductivity, such as zero electrical resistance, should be considered strong experiemental evidence of superconductivity.  In this work we demonstrated how to distinguish erromagnetic from superconducting signatures, further clarifying this identification proceedure.

\section*{Acknowledgements}
We thank Tej Nath Lamichhane for help in synthesis of Ce$_{3-x}$Mg$_x$Co$_9$ polycrystalline sample. S.L.B. thanks Vasily  Minkov for fruitful discussions. P.C.C. acknowledges past work by Francisco Urbano Ciervo as motivation for this work. This work was supported by the U.S. Department of Energy, Office of Science, Basic Energy Sciences, Materials Sciences and Engineering Division. The Ames National Laboratory is operated for the U.S. Department of Energy by Iowa State University under Contract No. DE-AC02-07CH11358. M.X. and W.X. at Michigan State University were supported by the U.S. Department of Energy, Division of Basic Energy Sciences under Contract DE-SC0023648. Single crystal growth of MnBi$_8$Te$_{13}$ (C.H., N.N.) was supported by the U.S. Department of Energy, Office of Science, Office of Basic Energy Sciences under Award Number DE-SC0021117.

\section*{Data availability}
The data that supports the findings of this article are openly available. \cite{dat26a}

\clearpage
\clearpage
\clearpage
\clearpage
\clearpage
\clearpage
\clearpage
\clearpage
\clearpage

\clearpage

\appendix

\section{Additional magnetization data for MnBi$_8$Te$_{13}$}
\label{A}
\setcounter{figure}{0} \renewcommand{\thefigure}{A\arabic{figure}} 

\begin{figure*}
  \includegraphics[width=0.8\textwidth]{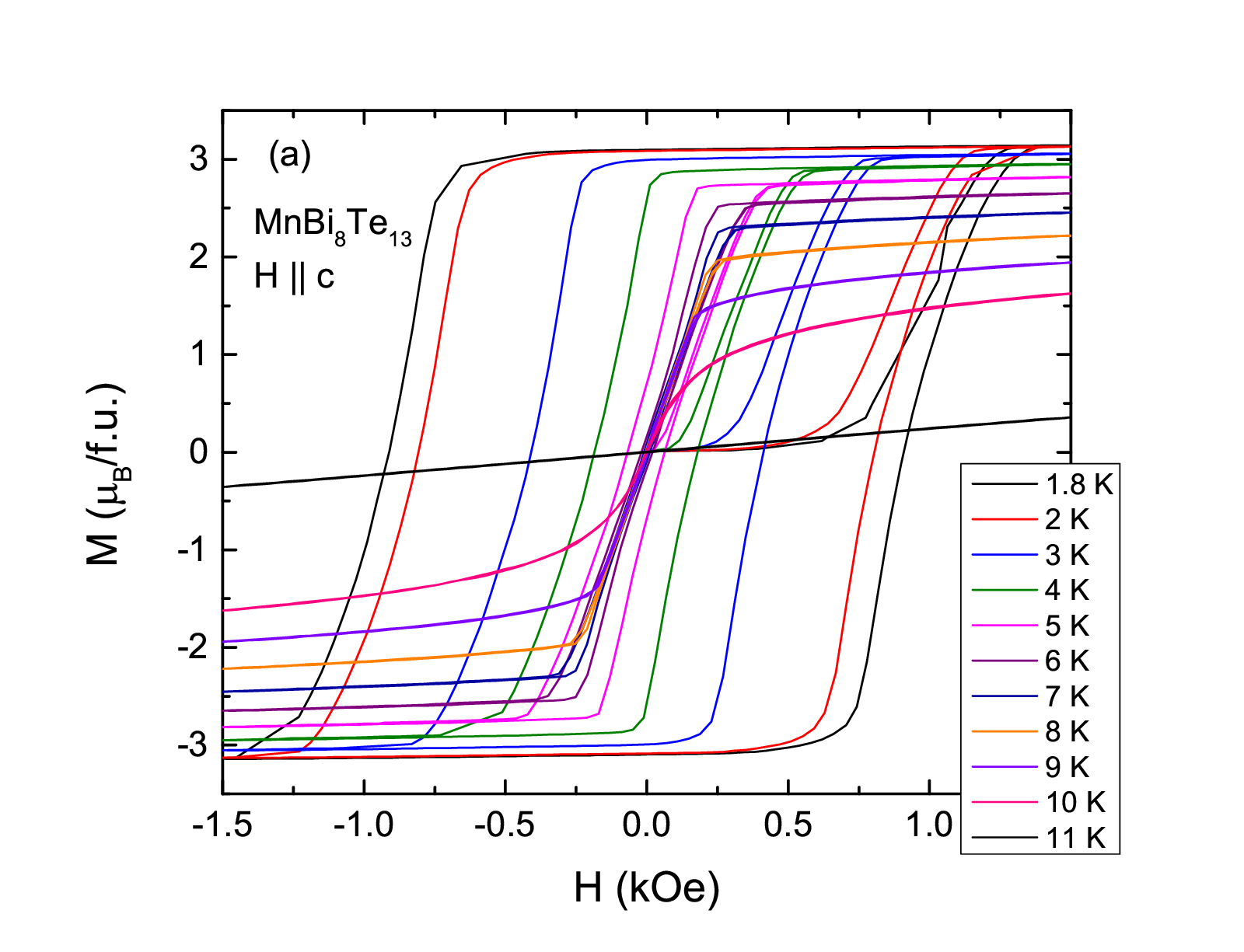}
  \includegraphics[width=0.8\textwidth]{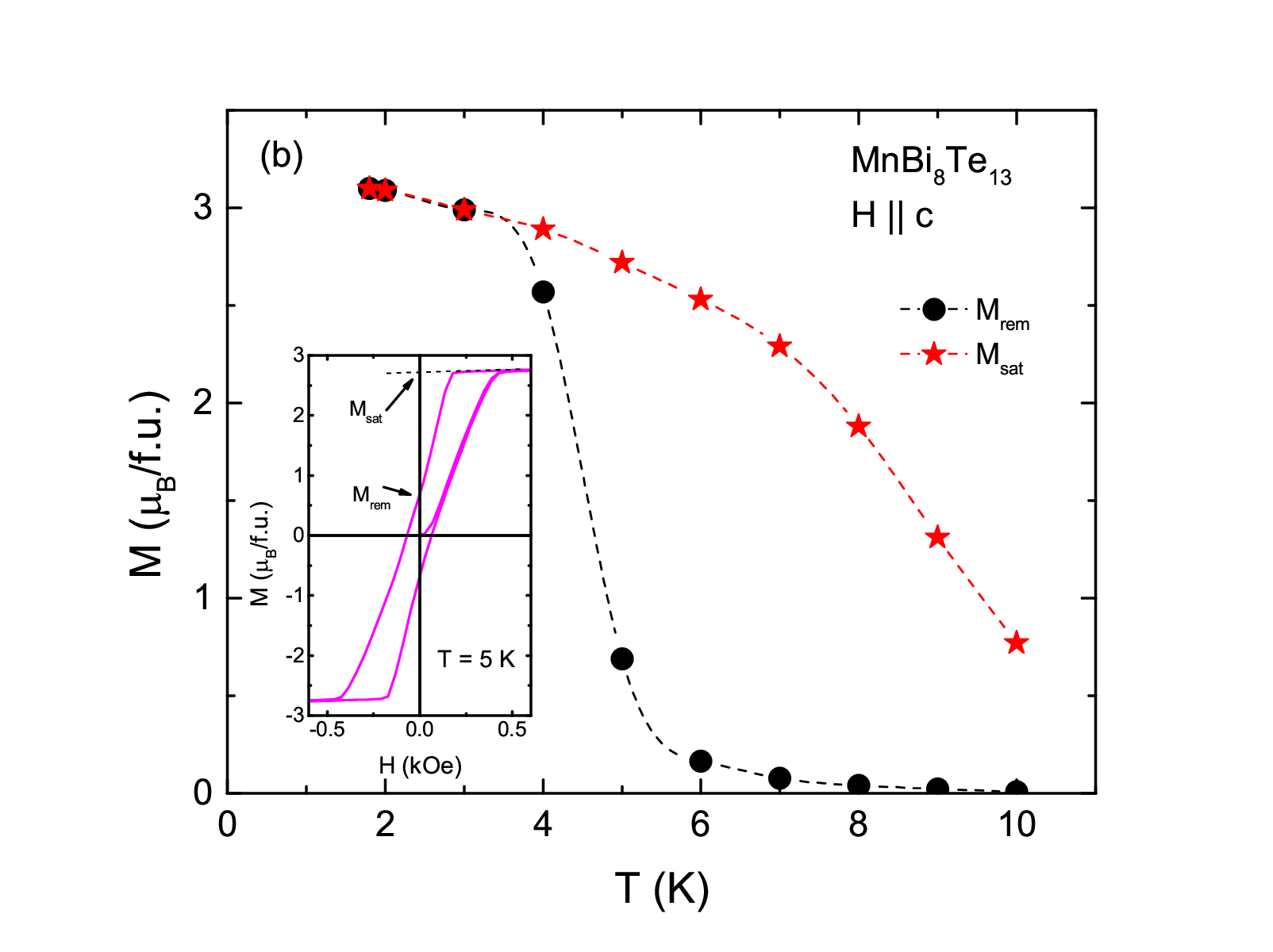}
\caption{(Color online) (a) As measured five-quadrants magnetization loops ($H \| c$) for MnBi$_8$Te$_{13}$. (b) Temperature dependence of remanent magnetization ($M_{rem}$) and saturation magnetization ($M_{sat}$) for MnBi$_8$Te$_{13}$ obtained from the magnetization loops in panel (a). Inset: the loop at $T = 5$~K with definitions of $M_{rem}$ and $M_{sat}$ shown by arrows.}
\label{FA1}       
\end{figure*}

Fig. \ref{FA1}(a) shows as measured, without demagnetizing field skewing correction, \cite{sko99a} five quadrant magnetization loops ($H \| c$) for MnBi$_8$Te$_{13}$ measured at different temperatures. These data are consistent witth a subset of loops reported in Ref. \cite{huc22a}. Based on these loops, temperature dependence of of remanent magnetization ($M_{rem}$) and saturation magnetization ($M_{sat}$) was inferred (Fig. \ref{FA1}(b), criteria are shown in the inset). So inferred $M_{rem}(T)$ and $M_{sat}(T)$ are significantly different already at 4~K ($\sim 0.4 T_C$) due to demagnetization field skewing anf rather narrow hysteresis loops.

\clearpage

\end{document}